\documentclass[prl,aps,superscriptaddress,showpacs,twocolumn]{revtex4-1}
\usepackage{graphicx,amsfonts,bm,amsmath}
\newcommand{\zt}{\tilde z}
\newcommand{\tit}{\tilde t}

\newcommand{\omt}{\tilde \omega}

\begin{document}

\author{Anna Grodecka-Grad}
\email{anna.grodecka-grad@nbi.dk}
\affiliation{QUANTOP, Danish National Research Foundation Center for Quantum Optics,
Niels Bohr Institute, University of Copenhagen, DK-2100 Copenhagen \O, Denmark}
\author{Emil Zeuthen}
\affiliation{QUANTOP, Danish National Research Foundation Center for Quantum Optics,
Niels Bohr Institute, University of Copenhagen, DK-2100 Copenhagen \O, Denmark}
\author{Anders S. S{\o}rensen}
\affiliation{QUANTOP, Danish National Research Foundation Center for Quantum Optics,
Niels Bohr Institute, University of Copenhagen, DK-2100 Copenhagen \O, Denmark}

\title{High-Capacity Spatial Multimode Quantum Memories Based on Atomic Ensembles}

\begin{abstract}
We study spatial multimode quantum memories based on light storage in extended ensembles of $\Lambda$-type atoms.
We show that such quantum light-matter interfaces allow for highly efficient storage of many spatial modes.
In particular, forward operating memories possess excellent scaling with the important physical parameters:
quadratic scaling with the Fresnel number and even cubic with the optical depth of the atomic ensemble. 
Thus, the simultaneous use of both the longitudinal and transverse shape of the stored spin wave modes
constitutes a valuable and so far overlooked resource for multimode quantum memories. 
\end{abstract}

\maketitle

\textit{Introduction}.---Photons are ideal candidates for carrying quantum information. 
In order to store and process the information, a quantum storage medium is, however, needed. 
To achieve this, one needs to establish a controllable and efficient light-matter interface
that will store light as a stationary excitation in a medium while preserving quantum correlations.
Quantum memories have already been demonstrated in a number of experiments based on atomic ensembles,
see e.g. \cite{Hammerer2010,Julsgaard2004,Novikova2007,Firstenberg2009,Choi2010,Reim2010,Hosseini2011},
as well as solid state systems, see e.g. \cite{Afzelius2010,Bonarota2011,Saglamyurek2011}.
Most of the realized memories support only a single mode, but it is highly desirable to be capable of storing as many modes as possible as this will increase the speed of quantum communication and facilitate quantum computation~\cite{Simon2007,Collins2007,Moelmer2008}.
To this end, several protocols have proposed exploiting various degrees of freedom to achieve multimode operation: spatial \cite{vasilyev2008} or directional \cite{surmacz2008,Moelmer2008} modes as well as frequency-multiplexing in the context of controlled reversible inhomogeneous broadening (CRIB) \cite{kraus2006,tittel2009} and time-binning with atomic frequency combs \cite{Afzelius2009}. The latter has been successfully realized experimentally with the storage and retrieval of four temporal modes~\cite{DeRiedmatten2008}. In addition, experimental realizations of memory qubits involving two co-existing spatial modes have been reported \cite{Inoue2006, Yuan2008}.
These results point toward promising applications of multimode quantum memories, but a full assessment of the potential of these requires an evaluation of the achievable memory capacity. Until now, this has only been performed in the one-dimensional (1D) case~\cite{Nunn2008}.

Here, we study the full capacity of the additional resource given by the spatial extent of atomic ensembles.
We show that combining the longitudinal and transverse degrees of freedom allows for highly efficient storage 
of many spatial light modes resulting in capacities higher than previously expected. 
The number of modes one can store with high efficiency depends on the choice of the direction of retrieval relative to that of the storage process.
We demonstrate that  forward operating memories, with the retrieved light traveling in the direction of the input signal,
provide an excellent multimode memory resource. 
It has a remarkable scaling with the important physical parameters:
the peak optical depth $d_0$ and the Fresnel number of the atomic ensemble~$F$. For broad ensembles ($F\gg 1$) each transverse mode can be described by the aforementioned 1D theory, which predicts the longitudinal mode capacity for backward retrieval to scale with $\sqrt{d_0}$ for Raman memories
as well as for protocols based on electromagnetically induced transparency (EIT) and with $d_0$ for CRIB protocols~\cite{Nunn2008,RemarkNunn}; in this Letter, we will consider the former two.
The dependence of the capacity on the Fresnel number has only been roughly estimated in Ref.~\cite{Golubeva2011} to be the number of transverse modes $\sim F^2$ for forward retrieval and $\sim F$ for backward retrieval. From the 1D calculations one would na\"{i}vely estimate the 3D capacity to be given by the number of transverse modes times the longitudinal capacity for each mode resulting in scalings of $F^2\sqrt{d_0}$ and $F\sqrt{d_0}$ for forward and backward retrieval, respectively.
Here, we show by direct calculation that the simultaneous use of the transverse and longitudinal shape of the stored spin wave mode leads to quantum memories with capacities scaling as $F^2 d_0^3$ for the forward direction. This is a much stronger scaling resulting in significantly higher memory capacities and, thus, far more promising forward operating memories than one would expect from previous work.

For comparison, we also study the backward operating spatial memory, with the retrieved light traveling in the opposite direction of the input light. Contrary to what is seen in the 1D limit \cite{Nunn2008}, backward operation generally possesses lower capacities, but we show that it can also serve as a high capacity multimode memory although with a slightly less promising scaling with the physical parameters.

\begin{figure}[tb] 
\begin{center} 
\includegraphics[width=7cm]{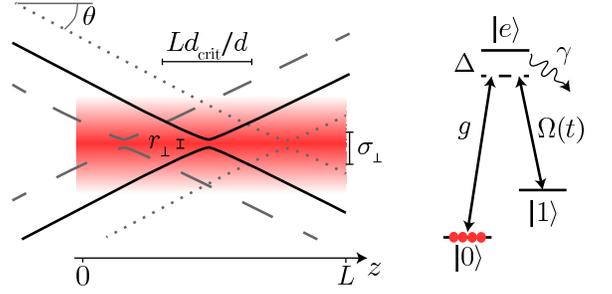}
\end{center} 
\caption{\label{fig:scheme}(color online). (left) Atomic ensemble of cylindrical symmetry with Gaussian density distribution of length $L$ and width $\sigma_\perp$ (grey/red cloud in the middle). Full lines: Multimode operation is achieved by focusing beams to a transverse size $r_\perp<\sigma_\perp$. For high optical depth it is an advantage to use light beams which are only focused inside the sample for a distance  $L d_{{\rm crit}}/d$ shorter than the length of the sample, so that one can store multiple modes along the axis (dashed and dotted lines, see text for details).   
(right) A schematic plot of the level scheme of the considered $\Lambda$-type atoms.}
\end{figure}

\textit{Model}.---In order to analyze the capacity of spatial multimode quantum memories, 
we use the three-dimensional theory for Raman and EIT quantum memories based on $\Lambda$-type atomic ensembles 
presented in Ref.~[\onlinecite{Zeuthen2011}].
There, it was shown that the crucial physical parameters determining the quality of the quantum memory are 
the optical depth $d_0$ and the Fresnel number of the atomic ensemble~$F$.
We consider a cylindrically symmetric atomic ensemble with a Gaussian distribution in the radial direction
$n(\rho) = n_0 \exp [-\rho^2/(2\sigma_\perp^2)]$ [see Fig.~\ref{fig:scheme} (left)],
where $n_0 = N_{\rm A}/(2\pi L \sigma_\perp^2)$, $N_{\rm A} $ is the number of atoms, 
$L$ is the length, and $\sigma_\perp \ll L$ describes the width of the cigar-shaped ensemble, e.g., corresponding to dipole trapped samples \cite{Kubasik2009,Appel2009}.
The density along the longitudinal $z$ axis has been assumed constant for simplicity. 
The geometry of the ensemble is described by its Fresnel number $F = \sigma_\perp^2/(\lambda_0 L)$,
where $\lambda_0$ is the wavelength of the quantum light.
The weak quantum field carries the quantum information to be stored into the atomic ensemble
and couples states $|0\rangle$ and $|e\rangle$ with coupling strength $g$ [see Fig.~\ref{fig:scheme} (right)]. 
States $|1\rangle$ and $|e\rangle$ are coupled by the strong classical control field, which sets the propagation direction of the retrieved light. 
$\Omega(t)$ is the Rabi frequency of the driving field and $\Delta$ denotes the detuning
from the excited state $|e\rangle$, which spontaneously decays at a rate $\gamma$.
The empty quantum memory is initialized to have all atoms in  state $|0\rangle$; storage is achieved by the absorption of photons from the light field, which entails the transfer of atoms from $|0\rangle$ to the state $|1\rangle$ via the intermediary state $|e\rangle$. More precisely, each photon is stored in a collective state of the ensemble represented by stationary
spin wave excitations described by $\hat{S} \sim \sum_i |0 \rangle_i\langle 1|$~\cite{Hammerer2010}. Working in the unsaturated limit, this atomic "spin" can be approximated as a set of harmonic oscillators.

For simplicity, we solve the three-dimensional problem of the multimode quantum light-matter interface
within the adiabatic approximation, where the excited state $|e\rangle$ is eliminated. To describe the transverse degrees of freedom we expand the slowly varying light field and spin wave operators on a complete set of transverse mode functions $\hat{a}(\vec{r},t)=\sum u_{mn}(\rho,\phi)\hat{a}_{mn}(z,t)$, $\hat{S}(\vec{r},t)=\sum u_{mn}(\rho,\phi)\hat{S}_{mn}(z,t)$. The light field and spin wave operators are then represented by vectors, $\vec{a}'(z,t)=\{\hat{a}_{mn}(z,t)\}$, $\vec{S}'(z,t)=\{\hat{S}_{mn}(z,t)\}$, containing a set of harmonic oscillator annihilation operators obeying 1D equations.
These equations of motion within the paraxial approximation (in the co-moving frame $t' = t-z/c$) read~\cite{Zeuthen2011}
\begin{eqnarray}\label{eq:adp}
  \frac{d}{d\zt} \vec a(\zt,\tit) & = & \left(-\frac{i \vec k_\perp^2 \sigma_\perp^2}{4\pi F} - \frac{\frac{1}{4}d_0}{\frac{1}{2}
+i\tilde \Delta} \mathbb{B}^2 \right)\vec a(\zt,\tit) \\ \nonumber
&& -\frac{\frac{1}{4}\sqrt{d_0}\tilde \Omega(\tit)}{\frac{1}{2}+i\tilde \Delta}\mathbb{B} \vec S(\zt,\tit), \\ \nonumber
\frac{d}{d\tit} \vec S(\zt,\tit) & = & -\frac{\frac{1}{4}|\tilde \Omega(\tit)|^2}{\frac{1}{2}+i\tilde \Delta} \vec S(\zt,\tit) 
-\frac{\frac{1}{4}\sqrt{d_0}\tilde \Omega^*(\tit)}{\frac{1}{2}+i\tilde \Delta} \mathbb{B} \vec a(\zt,\tit).
\end{eqnarray}
Here, we have introduced the dimensionless time $\tit = \gamma t'$, 
detuning $\tilde \Delta = \Delta /\gamma$, position $\zt = z/L$, and Rabi frequency $\tilde \Omega(\tit) = \Omega(\tit) /\gamma$. 
The peak optical depth $d_0 = 4Ln_0|g|^{2}/\gamma$ quantifies the absorption of resonant light 
in the absence of the control field $\tilde \Omega(\tit)$. We omit here the quantum noise since it is not needed for 
calculating the efficiency~\cite{Gorshkov2007a}.
Due to the sample symmetry as well as for numerical reasons, we have chosen a set of Bessel beams
indexed by $n$ and the azimuthal quantum number $m$. Furthermore, working in the paraxial regime and assuming a monochromatic signal, we have approximated $k_{||,mn} - k \approx - k^2_{\perp,mn}/(2k) \approx - k^2_{\perp,mn}/(2k_0)$, which allows us to account for the different transverse wave numbers of the modes through the term in Eq. (\ref{eq:adp}) containing the diagonal matrix $\vec{k}^2_\perp \equiv \{k^2_{\perp,mn}\}$.
The last terms in the above equations of motion describe the coherent interaction between light and matter,
which is quantified by the optical depth, Rabi frequency, detuning, and the matrix 
$B_{mn,m'n'} = \int d^{2} \vec r_{\perp} u_{mn}^{*} (\vec r_{\perp}) u_{m'n'} (\vec r_{\perp}) n(\vec r_\perp)/n_0$ describing the coupling between modes. The fact that this coupling matrix is independent of the axial position $\tilde z$ renders the problem solvable in terms of a simple matrix exponential.

In order to solve the equations of motion~(\ref{eq:adp}), we  Laplace transform in time
${\cal L}\{g(t)\}=\int_{0}^{\infty}e^{-\omega t}g(t)dt$, which allows us to eliminate the differential equation for the spin wave.
Here, we have assumed a constant driving field in space and time $\tilde \Omega$. 
In consequence, we can write the relation between the light modes and the spin wave 
(here for the forward operating memory) in the form of input-output beam splitter relations
\begin{eqnarray}
\vec{a}_{\text{out}}(\omt) & = &\int_{0}^{1}d\tilde{z}\; \mathbb{K}[\tilde \Omega,\omt,\tilde{z}]\; \vec{S}_0(\tilde{z}),\\
\vec{S}_0(\tilde{z}) &= &\frac{1}{2\pi i}\int_{-i\cdot\infty}^{i\cdot\infty}d\omt \mathbb{K}^{\text{T}}
[\tilde \Omega^{*},\omt,1-\tilde{z}]\vec{a}_{\text{in}}(\omt).
\end{eqnarray}
The transformation matrix $\mathbb{K}$ depends on frequency $\tilde \omega$, position $\zt$, and the physical parameters of the system
$\tilde \Omega, \tilde \Delta, F$, and $d_0$. 
The analytical expressions for the matrix as well as the beam splitter relations for the backward read-out are presented 
in detail in Ref.~[\onlinecite{Zeuthen2011}].
The efficiency of the quantum memory is the ratio between the number of outgoing and incoming light field excitations,
which, assuming a normalized incoming light mode, can be written as
\begin{eqnarray} \label{eq:eta}
\eta = \int_0^\infty d\tit |\vec a_{\rm out}(\tit)|^2 
\sim\iint d\nu d\nu' \vec a_{\text{in}}^\dag(\nu) \mathbb{M}[\nu,\nu'] \; \vec a_{\text{in}}(\nu').
\end{eqnarray}
Here, the kernel matrix $ \mathbb{M}$ gives full information about the relation between the input and output modes
for a given set of physical parameters.
After discretizing frequency and position, we diagonalize the large kernel matrix,
which for forward retrieval is
$\mathbb{M}[\nu,\nu'] \sim  \mathbb{K}^*[\tilde \Omega^*, \nu, 1-\zt]  \mathbb{K}^\dag[\tilde \Omega, \omt, \zt]$ 
$\mathbb{K}[\tilde \Omega, \omt, \zt']  \mathbb{K}^T[\tilde \Omega^*, \nu', 1-\zt']$.
A set of characteristic efficiencies is thereby obtained as the eigenvalues of this matrix together with the corresponding
set of incoming light modes $\vec a_{\text{in}}(\nu)$. 
The optimal incoming light mode to store into the atomic ensemble corresponds to the eigenvector with the highest eigenvalue,
which is the maximal efficiency of the memory.
The remaining eigenvectors correspond to orthogonal modes that can be stored with lower efficiencies.
We show in the following that, in general, there exist many light modes that give high efficiencies of the quantum memory. 

The figure of merit for multimode quantum memories is the capacity, which can be defined in at least two different ways.
Firstly, one can simply count the number of modes with an efficiency above a minimal value $\eta_{\rm min}$.
Secondly, a more sophisticated measure can be obtained from the quantum capacity of a Gaussian channel with efficiency~$\eta$,
$Q(\eta) = \max \{ 0, \log_2 |\eta| - \log_2 |1-\eta| \}$~\cite{Wolf2007}.
$Q(\eta)$ is the average number of qubits that can be perfectly stored and retrieved from
a particular mode with combined storage and retrieval efficiency $\eta$,
provided one has access to many copies of such memory and optimal encoding, decoding, and error correction of the stored information.
The capacity of the memory is obtained by summing the capacities for all modes; only modes stored and retrieved
with a combined efficiency above $\eta = 0.5$ contribute due to the no-cloning theorem~\cite{Wootters1982},
so that $C = \sum_{\eta_i > 0.5} Q(\eta_i)$. This capacity serves an important role as an upper limit for the memory. While the full capacity is hard to exploit experimentally, it provides a guide to the difficulty of achieving a certain capacity in practice since this increases near the theoretical optimum. We find below that the capacity is very high $(>10^3)$ for reasonable parameters ($F\approx 1$, $d_0\approx 100$). This is much higher than the number of modes which can be handled experimentally. Hence, spatial quantum memories provide an almost unlimited resource for multimode operation even without approaching the theoretical maximum.

\begin{figure}[tb]
\begin{center} 
\unitlength 1mm
{\resizebox{87mm}{!}{\includegraphics{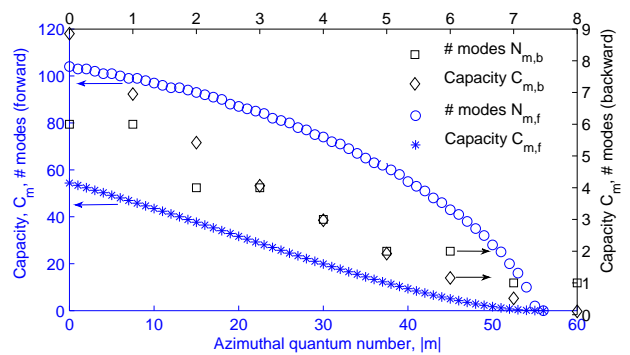}}}
\end{center} 
\caption{\label{fig:nm}(color online). The number of modes with efficiency $\eta > 0.5$ (squares and circles) and the capacity $C_m$ (diamonds and stars) for different values of the azimuthal quantum number $m$ 
for forward (blue, bottom and left axes) and backward (black, top and right axes) operating memory with $d_0 = 100$ and $F=1$.}
\end{figure}

\textit{Results}.---In order to gain insight into the multimode character 
we first calculate the capacity and the number of good orthogonal modes within subspaces
with a fixed azimuthal quantum number $m$, see Fig.~\ref{fig:nm}. 
All numerical results are obtained for resonant memories, $\tilde \Delta = 0$, the EIT case. The number of modes with high efficiency decreases with growing $|m|$
since light beams with $|m| > 0$ vanish to increasing degree toward the center of the atomic ensemble. 
Therefore, it is harder to focus the input light mode
into the dense center of the atomic cloud, leading to a decreased effective optical depth~$d_0$. 
One can see that for $d_0 = 100$ and $F=1$, the forward operating memory yields higher capacities than the backward one.
This can be explained by the fact that the driving light used for the backward retrieval reverses the longitudinal phase of the stored excitation
but cannot properly reverse its transverse profile except for a spin wave with a uniform transverse phase.
Thus in the case of any transverse phase gradient of the stored spin wave,
the irreversible transverse phase leads to unwanted diffraction effects~\cite{Zeuthen2011},
which in consequence reduce the efficiencies and the number of modes that can be stored in a backward operating memory.

\begin{figure}[tb]
\begin{center} 
\unitlength 1mm
{\resizebox{87mm}{!}{\includegraphics{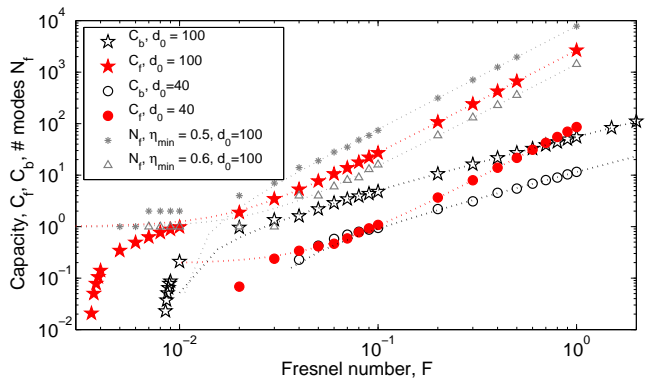}}}
\end{center} 
\caption{\label{fig:capacity}(color online). The capacity $C_{\mathrm f}$ for forward (red closed circles and stars)
and $C_{\mathrm b}$ backward (black open circles and stars) operating memories as functions of the Fresnel number~$F$
for $d_0 = 40$ (circles) and $d_0 = 100$ (stars). 
The number of modes for the forward memory $N_{\rm m,f}$ for $d_0 = 100$ with the threshold efficiency $\eta_{\min} = 0.5$ (light gray stars)
and $\eta_{\min} = 0.6$ (light gray triangles).
The dashed lines are  quadratic $C_{\rm f} \sim F^2$ and linear $C_{\rm b} \sim F$ fits for the forward and backward direction 
of the read-out, respectively.}
\end{figure}

The total capacity of forward operating quantum memories $C_{\mathrm f}$ is presented in Fig.~\ref{fig:capacity} (red closed circles and stars)
as a function of the Fresnel number of the atomic ensemble~$F$
for two values of the peak optical depth, $d_0 = 40$ and $100$.
In both cases, the total capacity $C_{\mathrm f}$ reaches high values and grows quadratically with the Fresnel number, $C_{\rm f} \sim F^2$.
To investigate whether the scaling is independent of the two capacity measures,
we also plot the number of modes $N_{\rm f}$ for $d_0 = 100$ for two values of the threshold efficiency  $\eta_{\min} = 0.5$ and $0.6$.
We see that the quadratic scaling with the Fresnel number $F$ is universal so that either of these may be used as the appropriate measure of the capacity.

The scaling with the Fresnel number depends, however, on the direction of the read-out. 
To compare the two operating modes of the memory, we also calculated the capacity of the backward memory (black open circles and stars), see Fig.~\ref{fig:capacity}.
We find that in this case, the capacity only scales linearly with the Fresnel number, but it still reaches high values.
Thus, even though the highest single mode efficiency for larger Fresnel numbers is achieved for backward operating memories~\cite{Zeuthen2011},
the number of good modes is larger for the forward memories for large $F$.

The scaling of the capacity can be explained by considering the diffraction of light in the atomic medium.
For forward operating memories, the divergence angle of the incoming light beam is $\theta \sim \lambda_0/r_\perp$,
where $r_\perp$ is the transverse waist of the stored stationary excitation. 
The mode can in this case be so focused that the maximum divergence angle becomes limited by the geometry of the ensemble:
$\tan \theta_{\rm max} \sim \theta_{\rm max} \propto \sigma_\perp/L$ (we ignore here a dependence on optical depth which will be included below).
From this we obtain that the minimal achievable waist of the stored excitation is $r_{\perp,\min} \propto \lambda_0 L/\sigma_\perp$.
In consequence, since the capacity of the memory is proportional to the ratio between the cross section area of the ensemble
and the minimal waist squared, this leads to a quadratic dependence 
on the Fresnel number, $C_{\rm f} \propto \sigma_\perp^2/r_{\perp,{\rm min}}^2 \sim F^2$. 
As mentioned above, in the case of backward operating memory, the problem of irreversible transverse phase arises~\cite{Zeuthen2011}. 
Therefore one cannot focus the light beam as strongly as in the case of the forward operating memory.
Requiring the phase to be constant across the transverse profile leads to $\theta_{\rm max} \propto  r_\perp/L$
and in consequence to the capacity of the memory $C_{\rm b} \propto   \sigma_\perp^2/r_{\perp,{\rm min}}^2 \sim F$,
showing linear dependence on~$F$.

\begin{figure}[tb]
\begin{center} 
\unitlength 1mm
{\resizebox{87mm}{!}{\includegraphics{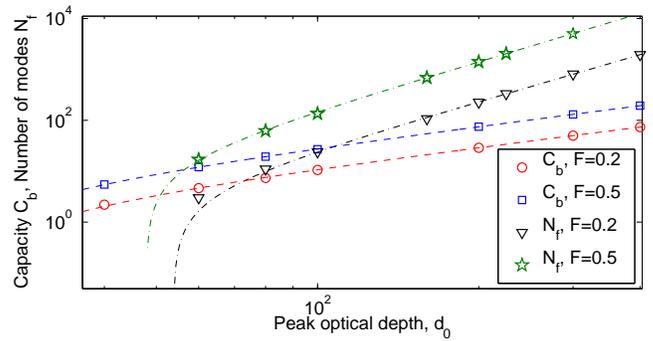}}}
\end{center} 
\caption{\label{fig:d0}(color online). The number of modes for the forward operating memory with efficiency $\eta \geq 0.65$
for two values of the Fresnel number $F=0.2$ (black triangles) and $F=0.5$ (green stars) and the capacity $C_{\rm b}$ of the backward operating memory 
for $F=0.2$ (red circles) and $F=0.5$ (blue squares). 
The dashed lines are the corresponding fits for the forward $N_{\rm f} \sim d_0^3$ and backward $C_{\rm b} \sim d_0^{3/2}$ direction of the read-out, respectively.}
\end{figure}

Above, we have presented the multimode character of the transverse degrees of freedom in the quantum memory
and its dependence on the Fresnel number of the atomic cloud. 
Now, we show that including simultaneously the transverse and longitudinal modes leads to much stronger scaling of the capacity with 
the optical depth $d_0$ than expected from considering these two degrees of freedom separately. 
We have optimized the full three-dimensional quantum memory and found that the capacity of the forward operating memory
has a promising cubic scaling with the optical depth $N_{\rm f} \sim d_0^3$, see Fig.~\ref{fig:d0}, much better than previously expected. (We use this measure since $C_{\rm f}$ is numerically cumbersome to calculate.)

This remarkable cubic scaling of the capacity of the forward operating memory with the optical depth $d_0$ can be understood by noting that for high optical depth it is not necessary to have the light beams confined within the transverse size of the sample for the entire length of the ensemble. Suppose that storage with a desired efficiency requires a certain critical optical depth $d_{\rm crit}$. This optical depth is achieved for a portion of the ensemble of length $Ld_{\rm crit}/d_0$. Hence it is only necessary to have the light transversely confined within the ensemble for this shorter distance allowing for larger divergence angles  $\theta_{\rm max}\sim \sigma_\perp d_0/Ld_{\rm crit}$, see Fig. \ref{fig:scheme}. (Alternatively, this relation can be understood by noting that with increasing optical depth along the axis, the optical depth also grows for beams incident at an angle. Beams can thus be incident at larger angles and still see an effective optical depth larger than $d_{\rm crit}$.) The larger divergence angle allows for stronger focusing of the beams down to a size $r_{\perp,\min} \sim d_{\rm crit}\lambda_0 L/\sigma_\perp d_0$ which increases the capacity. Furthermore since the storage of these tightly focused modes only involves a small portion of the sample, there will be  $d_0/d_{\rm crit}$ essentially independent storage media in the longitudinal direction, see Fig.~\ref{fig:scheme}. Combining the capacity of the transverse and longitudinal degrees of freedom we arrive at 
$N_{\rm f} \sim (\sigma_\perp/r_{\perp,\min})^2 d_0/d_{\rm crit} \sim F^2 d_0^3/d_{\rm crit}^3$. From an experimental perspective, this capacity reflects that the ensemble allow for storage of essentially any optical beam which can be focused into the ensemble and has a divergence angle less than $\theta_{\rm max}$. All of these modes are stored simultaneously using only a single control field, but the ideal temporal shape may be different for different modes and the exact set of spatio-temporal modes to use in a given experiment should be optimized given the experimental constraints. 
For comparison, we also provide the results for
backward operating memories, see Fig.~\ref{fig:d0}.
Here, the scaling is less promising $C_{\rm b} \sim d_0^{3/2}$, but still high values of the capacity are achievable. 

\textit{Conclusion}.---We have calculated the capacity of spatial quantum memories based on $\Lambda$-type, 
cigar-shaped atomic ensembles and thereby shown that they allow for storage of many light modes
and exhibit a remarkable scaling with the important physical parameters.
For memories operated in the forward direction, the capacity scales 
quadratically with the Fresnel number $F$ and cubically with the optical depth of the atomic ensemble $d_0$,
which is much better than previously expected~\cite{Nunn2008}.
These results reveal that the transverse degrees of freedom combined with the longitudinal ones
constitute a valuable resource for multimode quantum memories with excellent capacities. 
These results can be directly used in current experiments with extended ensembles of $\Lambda$-type atoms. 

This work was supported by HIDEAS (FP7-ICT-221906) and EMALI (MRTN-CT-2006-035369).
We thank A. Griesmaier, F. Kaminski, N.\,S. Kampel, J.\,H. M{\"u}ller, E.\,S. Polzik, and M.\,M. Wolf for fruitful discussions.~A.\,G.-G. and E.\,Z. contributed equally to this work.

\bibliographystyle{prsty}
\bibliography{longQM}

\end{document}